\documentclass{article}

\usepackage[dvips]{epsfig}

\date{February 2000}

\begin{document}

\title{\bf Lump scattering on the torus}

\author{RJ Cova\thanks{rcova@luz.ve} \\ Departamento de F\'{\i}sica
F.E.C. \\ La Universidad del Zulia \\
Apartado 15332 \\ Maracaibo 4005-A \\ Venezuela}

\maketitle


\abstract{Head-on collisions between two solitons in the pure $CP^1$ model on
a flat torus are investigated via numerical simulations. The charge-two
lumps, written out in terms of Weierstrass' elliptic $\wp$-function, are
found to scatter at 90$^{\circ}$. The phenomenon of singularity formation is
also seen. 

\section{Introduction}

The non-linear $O(3)$ sigma or $CP^1$ model in three-dimensional space-time is
a rich industry of research, both for its condensed matter applications and as
a simple field theory possessing topological solitons. The energy density
associated with the solitons are lumps localised in space.  The model also
appears as a low dimensional analogue of non-abelian gauge field theories in
(3+1) dimensions, an example being the increasingly popular Skyrme model of
nuclear physics. In pure mathematics the $CP^1$ solitons are known as harmonic
maps, by itself a long established area of research. 

The classical (2+0)-dimensional $CP^1$ model on the extended plane or Riemann
sphere \( \Re_{2} \cup \{\infty\} \simeq S_2 \), where the solitons are
harmonic maps \( S_2 \mapsto S_2 \)  given by any complex holomorphic
function, has been amply discussed in the literature \cite{lit1,lit2}.
The full time-dependent model [(2+1) dimensions] is not integrable, so
numerical simulations are needed for studying its dynamics.  Regarding
collisions, it is well known that the lumps on the extended plane scatter-off
at $90^{\circ}$ with respect to the initial direction of motion in the
centre-of-mass frame \cite{wjz}. Due to the conformal invariance of the planar
model, the lumps are unstable in the sense that they can shrink indefinitely,
leading to singularity formation in finite time \cite{leese1}. This
instability is cured by supplementing the $O(3)$ lagrangian with a Skyrme-like
and a potential-like term \cite{helv95}.

More recently, attention has also been paid to the $CP^1$ model with square
periodic boundary conditions, where the solitons are harmonic maps \( T_2
\mapsto S_2 \) ($T_2$ a flat torus) given by elliptic functions. Physically,
this approach looks more attractive than the one on $S_2$ because the system
is located in a finite volume from the outset. Mathematically, the model on 
$T_2$ has the formal advantage of removing the problem confronted in the
extended plane, whose non-compactness brings about difficulties to defining
the metric on the moduli space of static soliton solutions.

In reference \cite{non} we tackled the toroidal model by expressing the
soliton fields in terms of Weierstrass' $\sigma$-function. We found several
properties not seen in the usual planar theory, principally that there are no
single-soliton solutions on the torus and, in the Skyrme version of the
model, that there is no critical speed below which the skyrmions do not 
scatter at 90$^{\circ}$.

Our anterior paper \cite{split} resorted to the Weierstrass' elliptic
$\wp$-funcion to numerically study the $CP^1$ model, both in its original
and Skyrme versions, for systems with no initial speed. Novel features were
unveiled in the topological charge two sector. Amongst them: $\bullet$ the
appearance of four energy chunks instead of two for certain configurations
initially situated in the diagonals of the lattice and $\bullet$ lump
splitting in the Skyrme case for systems initially located in the central
cross of the grid. 

Within the framework of the geodesic approximation, the elliptic
$\wp$-function was employed as well by Speight \cite{martin} to describe pure
$CP^1$ lumps. His analysis predicted that the solitons may shrink and form
singularities in finite time --as their siblings on $S_2$-- and that the
solitons should also scatter at 90$^{\circ}$.

In the present paper we report the results for $CP^1$ lump scattering
obtained by numerically simulating the initial-value problem given by the
Weierstrass' $\wp$-function in the original, unmodified $CP^1$ model. Our
results bear out the predictions referred to in the previous parragraph. 

The paper is arranged as follows: In the next section we lay out the $CP^1$
model with periodic boundary conditions. In section 3 the numerical procedure
is explained and in section 4 the scattering results are discussed. We close
the paper with some concluding remarks including suggestions for further
research. 

\section{The \boldmath{$CP^1$} model on the torus} 

Our model is defined by the lagrangian density
%
\begin{equation}
{\cal L}=
\frac{|\partial_t W|^{2}-2 |\partial_z W|^{2}-
2 |\partial_{\bar{z}} W|^{2}}{(1+|W|^{2})^{2}} 
\label{eq:lagrangian}
\end{equation}
where \( z=x+iy \; \in \; T_2 \); $\bar{z}$ is the complex conjugate of $z$. 
The complex field $W$ obeys the periodic boundary condition
%
%
\begin{equation}
W[z+(m+in)L]=W(z), \qquad \forall t,
\label{eq:boundary}
\end{equation}
where $m,n=0,1,2,...$ and $L$ is the size of the torus. 
The static soliton solutions are elliptic functions which may be
written as
%
%
\begin{equation}
W=\lambda \, \wp(z-a) + b, 
\qquad \lambda, a, b \; \in \;{\cal Z},
\label{eq:w}
\end{equation}
being $\wp(z)$ the elliptic function of Weierstrass. Within a 
\emph{fundamental cell} of vertices
$$
(0,0),\, (L,0),\, (L,L),\, (0,L),
$$
$\wp$ is expandable as \cite{goursat}
%
%
\begin{equation}
\wp(z)=z^{-2} + \xi_2 z^2 + \xi_3 z^4 + ... + \xi_j z^{2j-2} + ...,
\quad \xi_j \; \in \; \Re.
\label{eq:p}
\end{equation}
This function is of the second order (degree) and so (\ref{eq:w}) 
represents solitons of topological charge 2. As discussed in
\cite{non,split} a distinctive feature of the toroidal model is the absence
of analytical single-soliton solutions. This can be understood by recalling
that the simplest non-trivial elliptic function is of order 2. In the
language of differential geometry, the degree of the harmonic maps \( M
\mapsto S_2 \)  must be greater than \( genus(M) \), $M$ a compact and
orientable Riemann surface.  Since \( genus(T_2)=1 \) there are no
unit-degree maps on the torus. However, a periodic single-soliton
\emph{ansatz} was constructed in \cite{non} using the equation
(\ref{eq:wsigma}) below with $\kappa$=1 and relaxing the accompanying
selection rule. Note that when $M=S_2$ we have the common model
with standard boundary conditions which do possess solitons in all homotopy
classes; this is because\linebreak \( genus(S_2)=0 \). 

In reference \cite{non} we computed the periodic solitons through
%
%
\begin{equation}
W=\prod_{j=1}^{\kappa} \frac{\sigma(z-a_{j})}{\sigma(z-b_{j})}, 
\qquad \sum_{j=1}^{\kappa} a_{j}=\sum_{j=1}^{\kappa} b_{j},
\label{eq:wsigma}
\end{equation}
with a subroutine that calculates $\sigma(z)$ numerically. 
Although $\sigma$ is pseudo-elliptic, the above ratio subject to the
summation of its zeros ($a_j$) being equal to the summation of its poles
($b_j$)  renders $W$ elliptic. In this paper we utilise a similar
subroutine and then reckon $\wp(z)$ using the formula \cite{goursat}: 
%
%
\begin{equation}
\wp(z)=-\frac{d^2}{dz^2}\ln[\sigma(z)].
\label{eq:psig}
\end{equation}
The Laurent expansion for $\sigma$ reads
%
\begin{equation}
\sigma(z)=\sum_{j=0}^{\infty}{c}_{j}z^{4j+1}, 
\qquad c_j \; \in \; \Re.
\label{eq:sig}
\end{equation}

Assisted by the useful properties of $\wp(z)$ :
%
%
\begin{equation}
\wp(z)=\wp(-z), \quad \wp(iz)=-\wp(z), \quad \wp(\bar{z})=\overline{\wp(z)},
\end{equation}
one readily deduces that $\wp(z)$ is real on the boundary and central cross
of the fundamental cell, and purely imaginary on the borders \cite{lawden}.

The \emph{static} (or potential) energy density $E$ associated with the
harmonic map $W$ may be read-off from the lagrangian (\ref{eq:lagrangian}).
Taking into account the identity \cite{lawden}
%
%
\begin{equation}
[\frac{d\wp(z-a)}{dz}]^2=4 \wp(z-a)[\wp(z-a)^2-\wp^2(L/2)],
\label{eq:id}
\end{equation}
we have
%
%
\begin{equation}
E=8 |\lambda|^2 \frac{|\wp(z-a)||\wp^{2}(z-a)-\wp^{2}(L/2)|}
                        {[1+|\lambda\wp(z-a)+b|^2]^2}.
\label{eq:eden}
\end{equation}
 
\section{Basic numerical procedure}
%
We treat configurations of the form (\ref{eq:w}) as the 
initial conditions for our time evolution, studied numerically. 
The time derivative of $W$ is calculated from the Lorentz-boosted
field. Our simulations run in the $\phi$-formulation of the
model, whose field equation 
%
%
 \begin{equation}
\partial^{2}_{t}\overrightarrow{\phi}=[-(\partial_{t}\overrightarrow{\phi})^{2}+
(\partial_{x}\overrightarrow{\phi})^{2}+
(\partial_{y}\overrightarrow{\phi})^{2}]\overrightarrow{\phi}+
\partial^{2}_{x}\overrightarrow{\phi}+\partial^{2}_{y}\overrightarrow{\phi}
  \label{eq:motioncp1}
\end{equation}
follows from the lagrangian density (\ref{eq:lagrangian}) with the help of 
%
%
\begin{equation} 
W=\frac{1-\phi_{3}}{\phi_{1}+i\phi_{2}}.
\label{eq:wphi}
\end{equation}
The real scalar field \( \overrightarrow{\phi}=(\phi_1,\phi_2,\phi_3) \)
satisfies \( \overrightarrow{\phi}.\overrightarrow{\phi}=1 \). Inverting
formula (\ref{eq:wphi}) entails
\begin{equation}
\overrightarrow{\phi}=(\frac{W+\bar{W}}{|W|^{2}+1},
i\frac{-W+\bar{W}}{|W|^{2}+1},\frac{|W|^{2}-1}{|W|^{2}+1}).
  \label{eq:projection}
\end{equation}

We compute the series (\ref{eq:sig}) up to the fifth term, the coefficients
$c_j$ being in our case negligibly small for $j \geq 6$. We employ the
fourth-order Runge-Kutta method and approximate the spatial derivatives by
finite differences. The laplacian is evaluated using the standard
nine-point formula and, to further check our results, a 13-point recipe is
also used. Our results show unsensitiveness to either technique, thus
confirming the reliability of our results. The discrete model evolves on a
200 $\times$ 200 periodic lattice ($n_x=n_y=200$) with spatial and time
steps $\delta x$=$\delta y$=0.02 and $\delta t$=0.005, respectively. The
size of our fundamental, toroidal network is $L=n_x \times \delta x=4$.

Unavoidable round-off errors gradually shift the fields away from the
constraint \( \overrightarrow{\phi}.\overrightarrow{\phi}=1 \). So we rescale
\( \overrightarrow{\phi} \rightarrow
\overrightarrow{\phi}/\sqrt{\overrightarrow{\phi}.\overrightarrow{\phi}} \) 
every few iterations.  Each time, just before the rescaling operation, we
evaluate the quantity \( \mu \equiv
\overrightarrow{\phi}.\overrightarrow{\phi} - 1 \)\, at each lattice point.
Treating the maximum of the absolute value of $\mu$ as a measure of the
numerical errors, we find that max$|\mu|$ $\approx$ 10$^{-10}$.  This
magnitude is useful as a guide to determine how reliable a given numerical
result is. Usage of an unsound numerical procedure in the Runge-Kutta
evolution shows itself as a rapid growth of max$|\mu|$; this also occurs, for
instance, when the unstable energy lumps become too spiky. 

\section{Results}
%
The initial conditions are then given by equation (\ref{eq:w})
$$
W=\lambda \, \wp(z-a) + b,
$$
where $\lambda$ is related to the size of the solitons, $b$ determines
their mutual separation and $a$ merely shifts the solution on the torus.
Calling the initial speed $v$, we boost the above field and take its time 
derivative:
%
%
\begin{equation}
\left.
\begin{array}{lllll}
W & \rightarrow & W(t)&=&\lambda \wp(z-a) + b(1-vt), \\
  &             & \partial_t W(t)&=&-bv,

\label{eq:wboosted}
\end{array}
\right.
\end{equation}
thus completely defining our initial-value problem.  Without loss of
generality we may take the values of these parameters according to numerical
convenience.
Let us set
$$
\lambda=1, \quad
a=(2.015, 2.015), \quad
b=1
$$ 
throughout. As depicted in figure \ref{fig:fig1}, the total energy density
corresponding to our soliton field has the form of two lumps sitting 
on the central cross of the cell, symmetrically around its centre.

Consider the situation where the lumps are sent towards one another with a
relative initial speed of $v=0.3$. We observe that the solitons gradually
expand as they approach each other; they collide at the centre of the nett and
coalesce into a ringish structure, where the solitons are no longer
distinguishable. Here they attain maximum expansion, {\em i.e.,} the peak of
the total energy density ($E_{max}$) reaches a minimum value. After this
process the lumps get narrower and narrower as they re-emerge at right angles
to their initial line of approach. They keep shrinking while moving away from
the centre, in opposite senses. Some time later they become so spiky that the
numerical code breaks down. This is the well-documented instability of the
$O(3)$ model in two dimensions, where the theory in conformally invariant. 
The foregoing events are illustrated in figure \ref{fig:fig2}.

The evolution of $E_{max}$ is shown in the upper half of figure
\ref{fig:fig3} for two values of $v$. The curves are qualitatively alike,
the life of the system being longer the smaller its initial velocity is. 

The kinetic energy remains very small all along except when the instability
takes over. This can be appreciated from figure \ref{fig:fig3} (bottom),
where the maximum value of the kinetic energy density ($K_{max}$) is plotted
{\em versus} time.  

Solitons with no initial velocity remain motionless with the passing of time
(see below, however, the particular case when $b=0$).  They too exhibit
shrinking and singularity formation, if at a slower rate than when $v\ne 0$;
the evolution of the total energy density qualitatively resembles that of
figure \ref{fig:fig3} (top).  As for the kinetic energy density, it stays very
small, at around $\approx 10^{-7}$ during the whole simulation (except when
the breadth of the lumps is comparable to the lattice spacing which leads to
the collapse of the numerics).

%
Particularly noteworthy is the case $b=0$.  The soliton field and
its energy distribution are, respectively from equations (\ref{eq:w})  and
(\ref{eq:eden}),
%
\begin{equation}
\left.
\begin{array}{lll}
W&=&\lambda \wp(z-a), \\ 
E&=&8 |\lambda|^2 \frac{|\wp(z-a)||\wp^{2}(z-a)-\wp^{2}(L/2)|}
                        {[1+|\lambda\wp(z-a)|^2]^2}.
\label{eq:wbzero}
\end{array}
\right.
\end{equation}

The global maxima of such energy density 
are located along the diagonals of the basic cell \cite{martin}, \emph{i.e.},
where $\wp(z)$ is purely imaginary.  Given that $E$ in (\ref{eq:wbzero}) is
invariant under \( \wp \rightarrow -\wp \), it follows from the evenness
 of $\wp$ that $E$ has at least for peaks on the diagonals of the fundamental
cell. In fact, it possesses four peaks (see figure \ref{fig:fig4}). This
situation in the topological index-two class is a remarkable feature of
$\wp$-solitons, going beyond the two-lump and annular structures found in the
$CP^1$ on $S_2$.

Now, the scattering problem when $b=0$ is especial: In order
to zoom the lumps towards each other one must boost $W$ according to
equation (\ref{eq:wboosted}), \emph{i.e.}, set \( b \rightarrow b(1-vt) \).
But this is only possible if $b \neq 0$.  And boosting the parameter $a$ in
$W$ is of no avail, for it simply shifts the system as a whole. 
Let us then consider systems of the form (\ref{eq:wbzero}) started off from
rest. It turns out that such configurations shrink quite slowly with time, as
illustrated in the upper graph of figure \ref{fig:fig5}.
From the bottom-right plot of figure \ref{fig:fig5} we observe 
that the solitons move away from the centre and back in along
the diagonals of the grid in
breather-like fashion. The kinetic waves emitted in the process evolve 
according to the bottom-left side of illustration number \ref{fig:fig5}.

Given similar set-ups, $CP^1$ lumps with periodic boundary conditions will
usually shrink at a slower rate than their relatives with standard boundary
conditions. This is because solitons in compact domains as $T_2$ always
coalesce somehow, and these are configurations with $E_{max}$ smaller than
when the solitons do not overlap --a situation achivable to a good
approximation in non-compact spaces like the extended plane. States with
$b=0$, which correspond to coincident structures that spread themselves all
over the cell (see picture \ref{fig:fig4}), have a small $E_{max}$
which evolves quasi-stably as depicted in figure \ref{fig:fig5}.  
Note that other than through the parameter $b$, we can always bring the
solitons to greater overlapping either by increasing the lump size ($\lambda$)
or by reducing the torus size ($L$). This is be useful if we want to
run a collision experiment with the same value of $b$ but different initial
values of $E_{max}$.

We point out that in reference \cite{split} our configurations for $b=0$
exhibited scattering in the Skyrme version of the model, even though the
initial speed was zero. We define our Skyrme model on the
torus by [compare with the lagrangian (\ref{eq:lagrangian})]

\begin{eqnarray}
{\cal L}&=&\frac{|\partial_t W|^{2}-2 |\partial_z W|^{2}-
2 |\partial_{\bar{z}} W|^{2}}{(1+|W|^{2})^{2}}
+ \nonumber \\
        & & \theta_{1}\frac{|\partial_z W|^{2}}{(1+|W|^{2})^{4}}
             (|\partial_{t} W|^{2}-|\partial _{z} W|^{2}).
\label{eq:skylagrangian}
\end{eqnarray}
In this scheme the four energy peaks move and collide along the diagonals and
scatter at right angles. Let us emphasise that the extra term in the
lagrangian (\ref{eq:skylagrangian}) stabilises the $CP^1$ lumps, \emph{i.e.},
prevents the formation of singularities.

As pointed out in the introduction, our main results for $\wp$-solitons
(singularity formation and scattering at 90$^{\circ}$)  conform with
prognostications made in the geodesic approximation treatment of the model
\cite{martin}. There it was proved that the moduli space of static 2-lump
solutions is geodesically incomplete and has finite diameter, leading to
infinite lump shrinkage. It was also proved in \cite{martin} that some
mathematical constraints oblige the lumps to lie either on the central
cross, the boundary or the diagonals of the lattice, implying that only
scattering at right angles should occur.   

\begin{figure*}[p]
\mbox{\epsfig{file=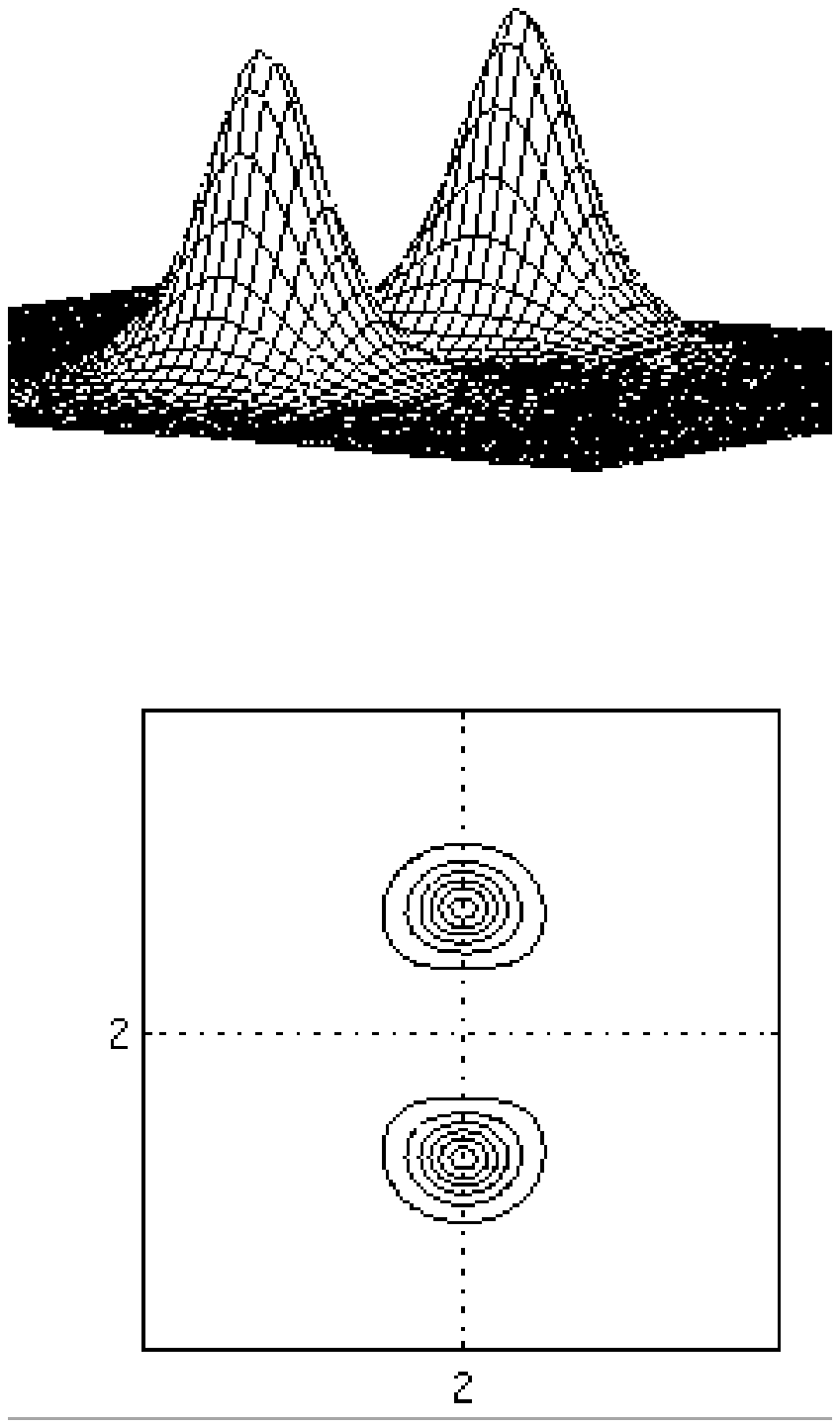}}
\caption{Three dimensional and contour pictures corresponding to
the total energy density of the solitons at $t=0$. The initial speed is
$v=0.3$.}
\label{fig:fig1}
\end{figure*}   

\begin{figure*}[p]
\mbox{\epsfig{file=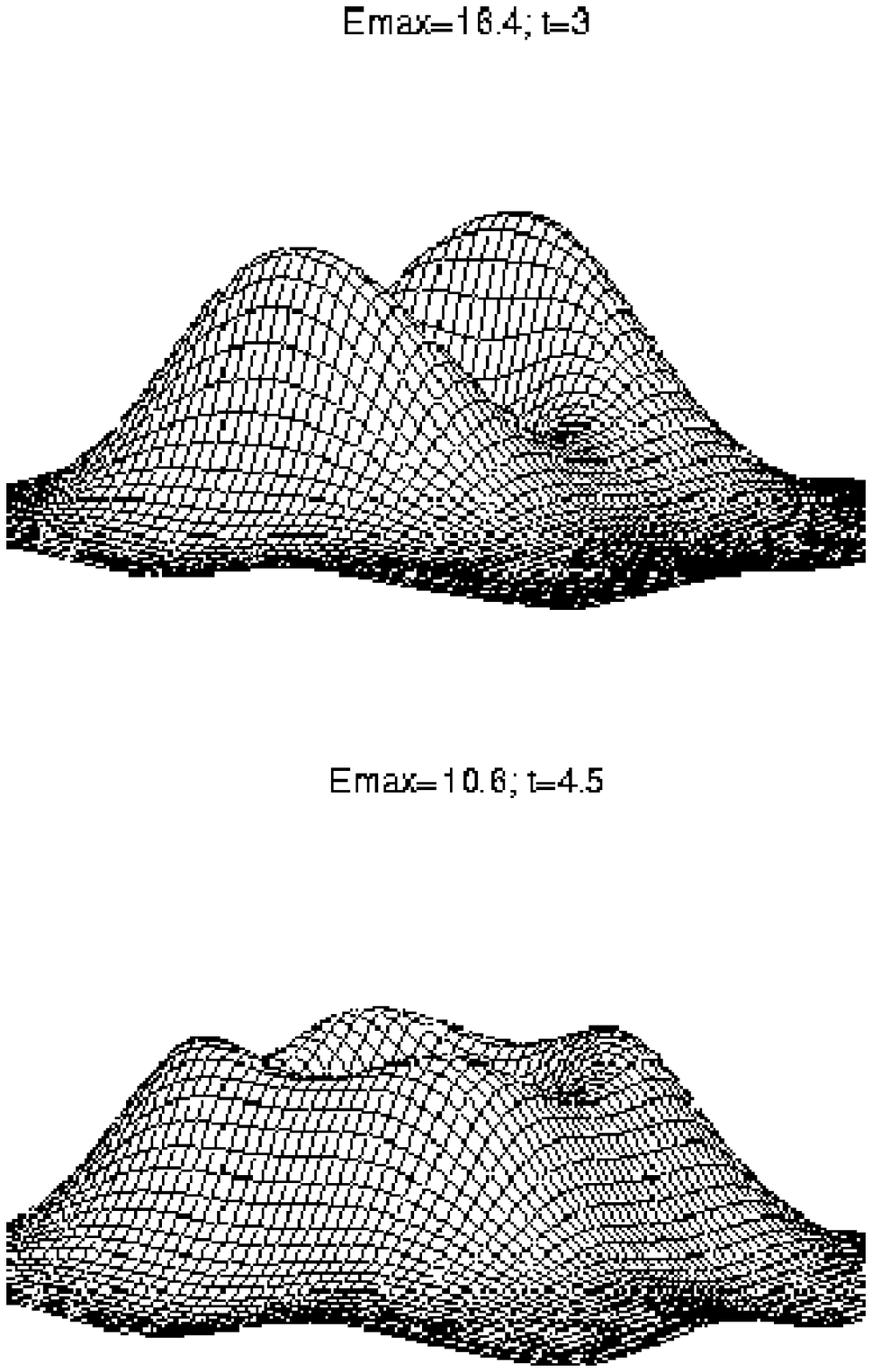}}
\caption{The lumps of figure \ref{fig:fig1} motion towards each other.
They collide and coalesce around the centre of the lattice, where they expand
maximally.}
\label{fig:fig2}
\end{figure*}

\begin{figure*}[p]
\mbox{\epsfig{file=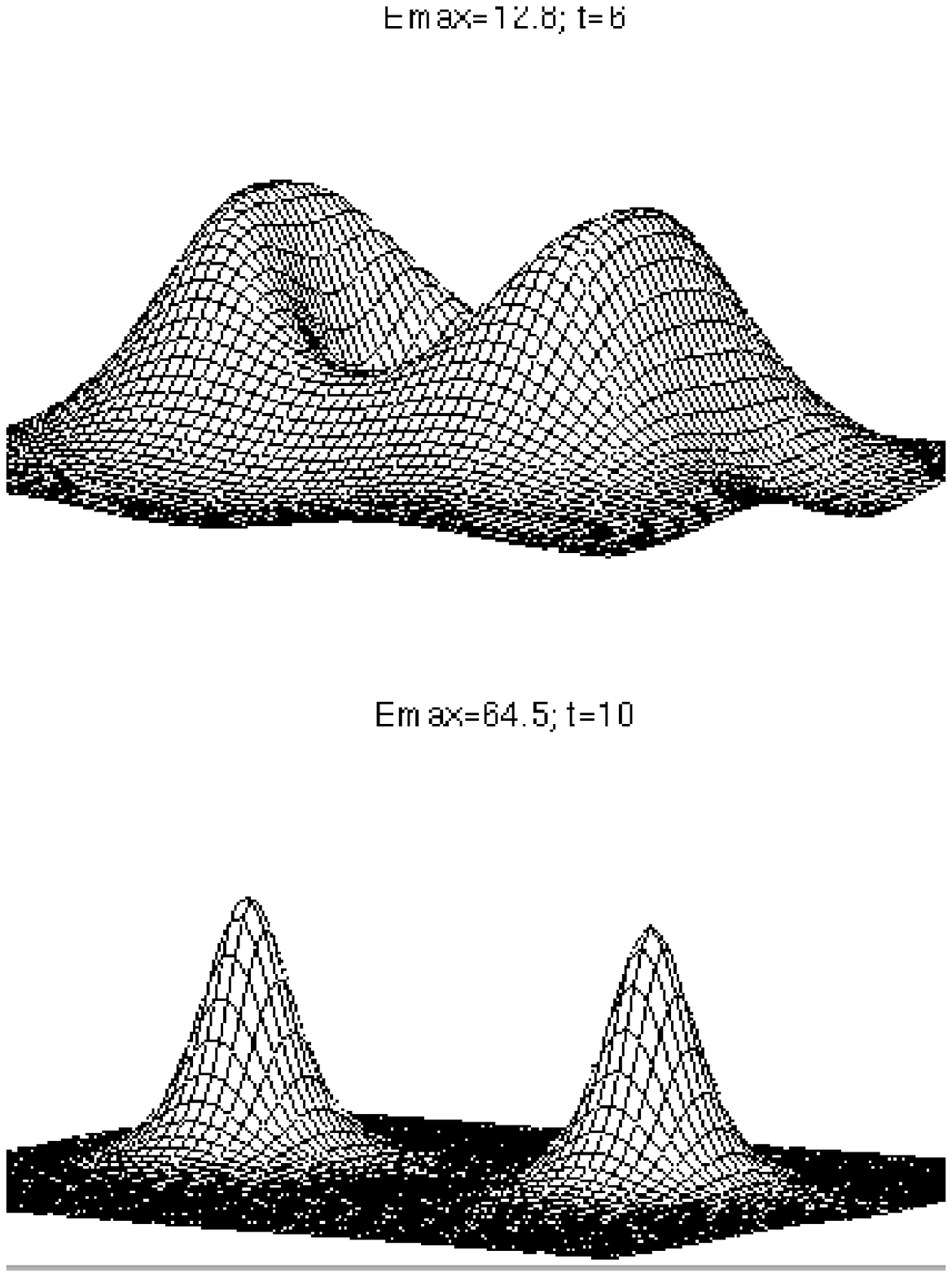}}
\begin{center}
\normalsize{Figure 2: Continued. The lumps scatter at
90$^{\circ}$ to the original direction of motion.}
\end{center}
\end{figure*}      

\begin{figure*}[p]
\mbox{\epsfig{file=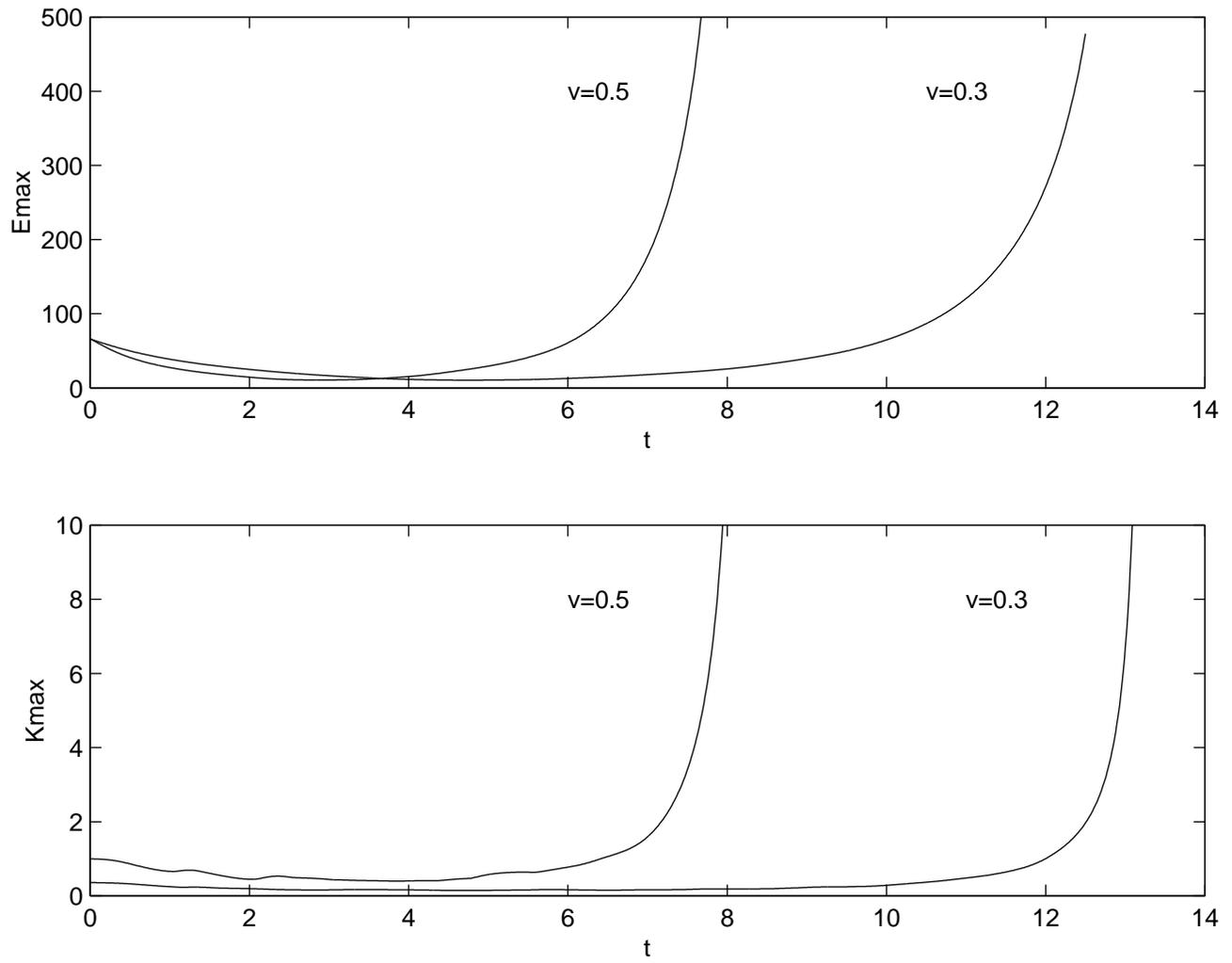}}
\caption{Evolution of the peak of the total energy density ($E_{max}$)
and the peak of the kinetic energy density ($K_{max}$) for two values of the
initial speed.}
\label{fig:fig3}
\end{figure*}        

\begin{figure*}[p]
\epsfig{file=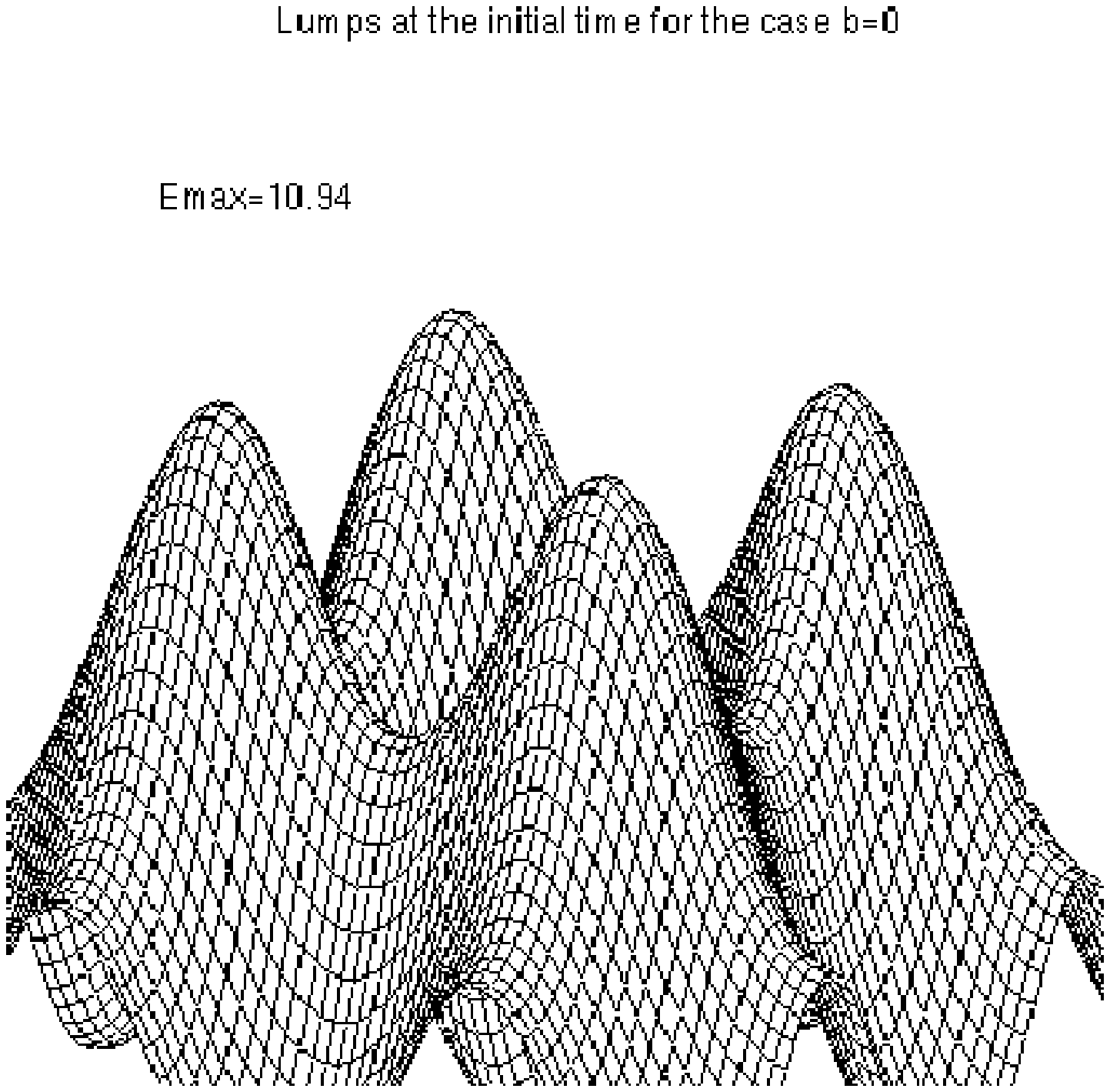}
\caption{Lumps at $t=0$ corresponding to the particular case $W=\wp(z-a)$. 
The presence of four lumps rather than two in the charge two 
sector is an alluring characteristic of $\wp$-solitons.}
\label{fig:fig4}
\end{figure*}

\begin{figure*}[p]
\epsfig{file=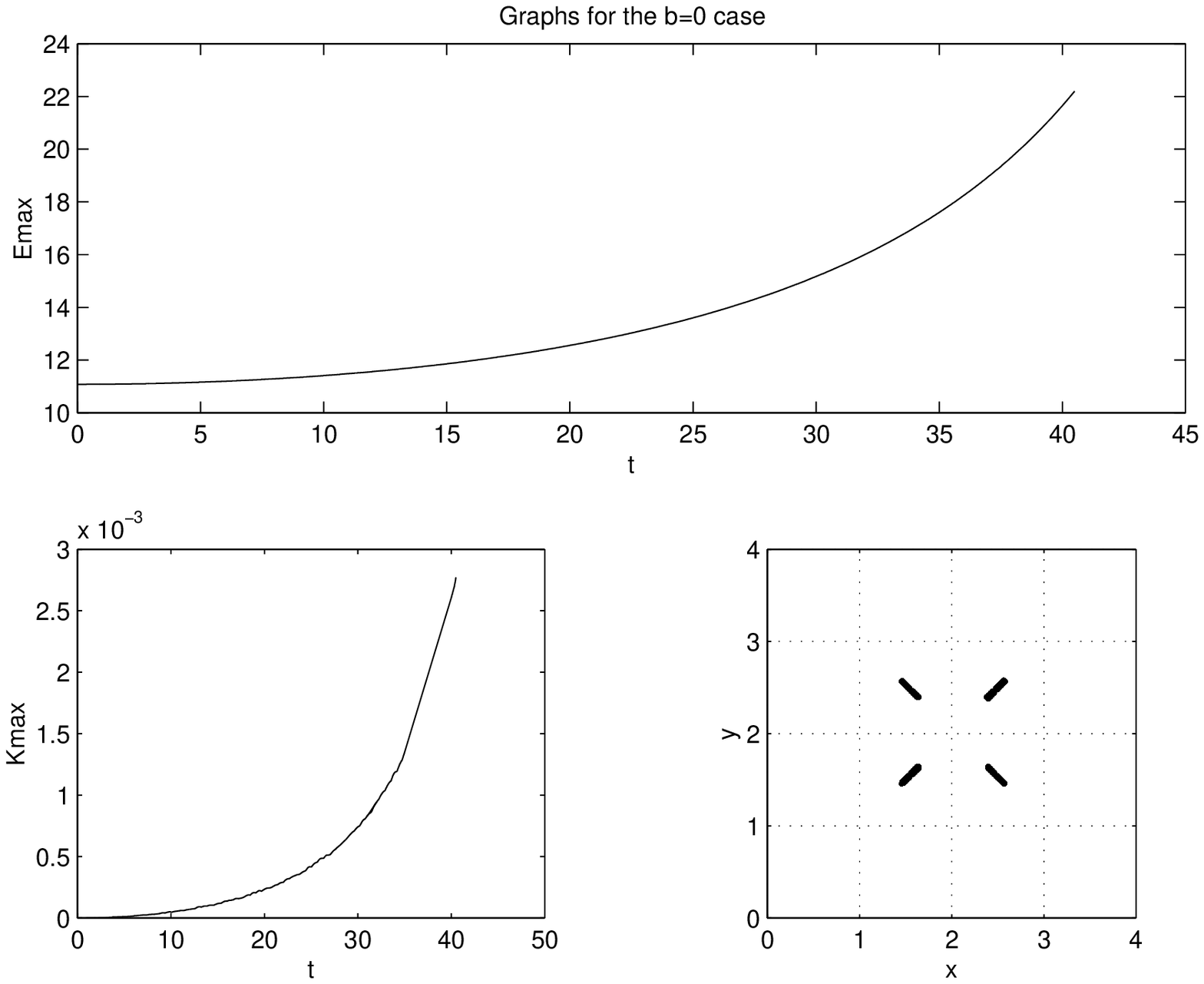}
\caption{Plots corresponding to $W=\wp(z-a)$. We observe a 
quasi-stable configuration where the lumps oscillate along the diagonals
of the network in breather-like manner. The initial velocity is zero.}
\label{fig:fig5}
\end{figure*}    

\section{Conclusions}

With the help of numerical simulations we have investigated some scattering
properties of the (2+1) dimensional nonlinear $O(3)$ or $CP^1$ model with
periodic boundary conditions, expressing the soliton solutions through
Weierstrass' elliptic $\wp$-function. Limiting ourselves to the pure version
of the model, we have observed that the lumps scatter off forming
$90^{\circ}$ with the initial direction of motion in the centre-of-mass
frame. During this process the lumps grow spiker as time elapses, eventually
breaking down the numerical code.  Lack of stability shows as well for lumps
with zero initial speed, but at a slower rate. The $O(3)$ instability may be 
understood from the symmetry of the model under dilation transformations. The
above  properties are basically the same as those known from the familiar
model on the Riemann sphere. But some other phenomena turn out to
be quite disimilar on $T_2$, \emph{e.g.}, the absence of unit-charge solitons
and the
presence of four lumps rather than two in the charge-two topological class.
The non-existence of single-soliton fields on $T_2$ is dictated by a
\underline{general} property stating that there are no elliptic functions of
order one. The four peaks in the charge-two sector stem from a notable
symmetry property \underline{specific} to the $\wp$-energy density when $b=0$
[see equation (\ref{eq:wbzero})]. A configuration of this type is not
observed when the fields are constructed out of the $\sigma$-function.

A natural extension of the present work is the study of
$\wp$-lumps with non-zero initial velocity in the Skyrme format. 
Skyrmions with such initial conditions have already shown a
thought-provoking scattering and splitting pattern under numerical
simulations.  Consequently, research on their evolution under boosting
seems compelling.  Further investigations may involve defining the solitons
in terms of other elliptic functions as those of Jacobi, and the
pseudo-elliptic $\theta$-functions. A classification of solitonic
properties on the torus can only be made after thorough consideration of
the above-mentioned functions. Let us remind that on $S_2$ no new traits
arise from casting the soliton configurations in different ways. In the
topological charge two class, for example, the following fields behave
qualitatively alike on the Riemann sphere:  $$ W= z^2,z^{-2},
\frac{(z-a)(z-b)}{(z-c)(z-d)}.  $$

It is a non-trivial problem to understand the mechanisms underlying the
complicated $CP^1$ dynamics. In the quest for such understanding, a study of
solitonic behaviour on compact manifolds, especially $T_2$, is certainly
quite an appealing program to pursue. 

\newpage
\Large \textbf{Acknowledgement}
\normalsize

The author is very grateful to WJ Zakrzewski for helpful conversations and
to the Mathematical Sciences Dept. of the University of Durham, England,
where part of the present work was carried out. The financial support of
{\em La Universidad del Zulia} through CONDES 218-99 is acknowledged with
big thanks. 


\end{document}